\documentclass[prl,reprint,amsmath,amssymb,showpacs]{revtex4-1}
\usepackage{graphicx}
\usepackage{hyperref}
\usepackage{color}

\begin{document}

\title{Giant slip at liquid-liquid interfaces using hydrophobic ball bearings}

\date{\today}

\author{Quentin Ehlinger}
\author{Laurent Joly}\email{laurent.joly@univ-lyon1.fr}
\author{Olivier \surname{Pierre-Louis}}\email{olivier.pierre-louis@univ-lyon1.fr}
\affiliation{Institut Lumi\`ere Mati\`ere, UMR5306 Universit\'e Lyon 1-CNRS, Universit\'e
de Lyon 69622 Villeurbanne, France}

\begin{abstract}
Liquid-gas-liquid interfaces stabilized by hydrophobic beads
behave as ball bearings under shear and exhibit giant slip.
Using a scaling analysis and Molecular Dynamics simulations we predict that, when the contact
angle $\theta$ between the beads and the liquid is large, the slip length 
diverges as $R\rho^{-1}(\pi-\theta)^{-3}$
where $R$ is the bead radius, and $\rho$ is the bead density.
\end{abstract}







\maketitle

\begin{figure}
\includegraphics[width=0.8\linewidth]{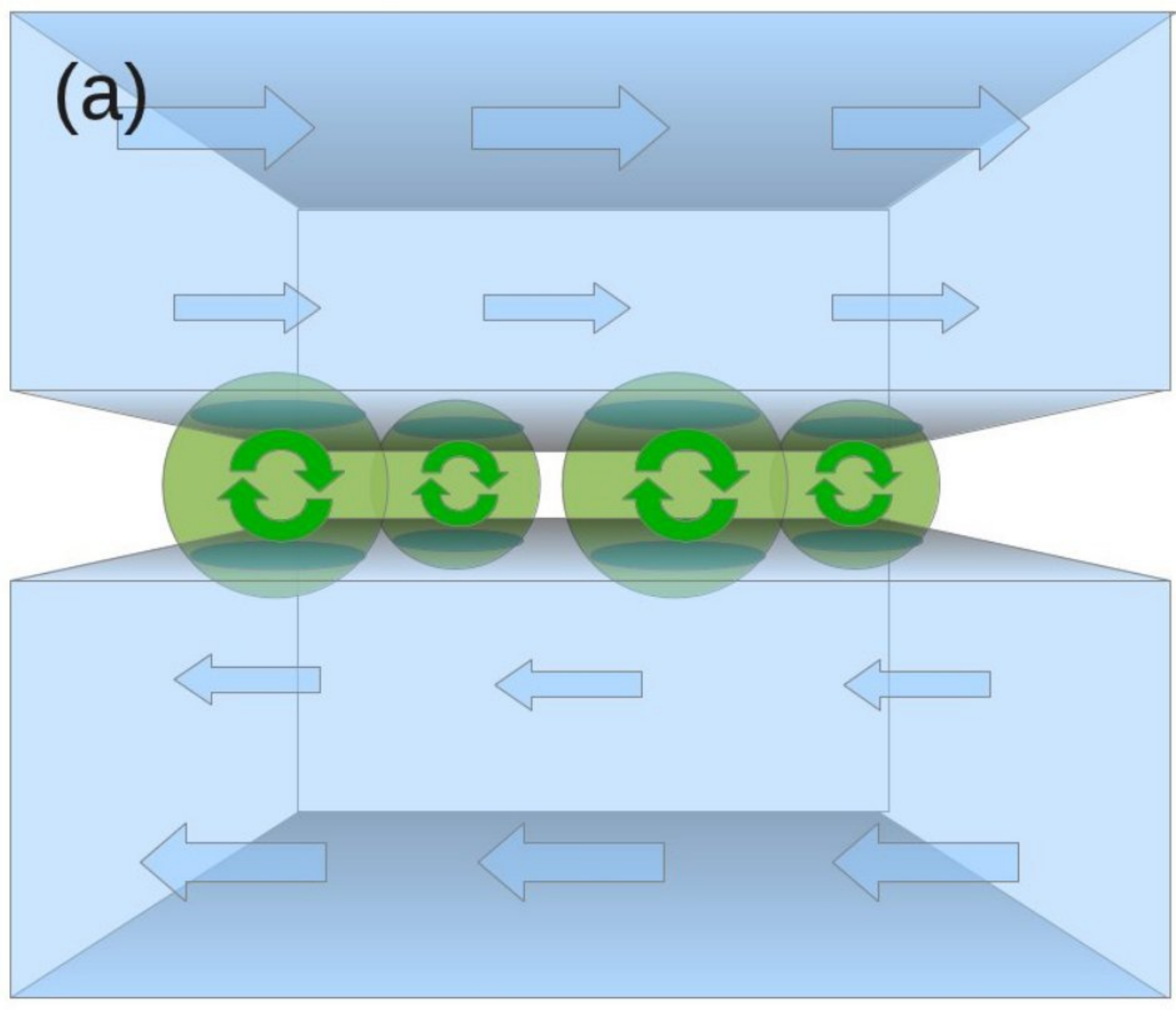}
\includegraphics[width=0.8\linewidth]{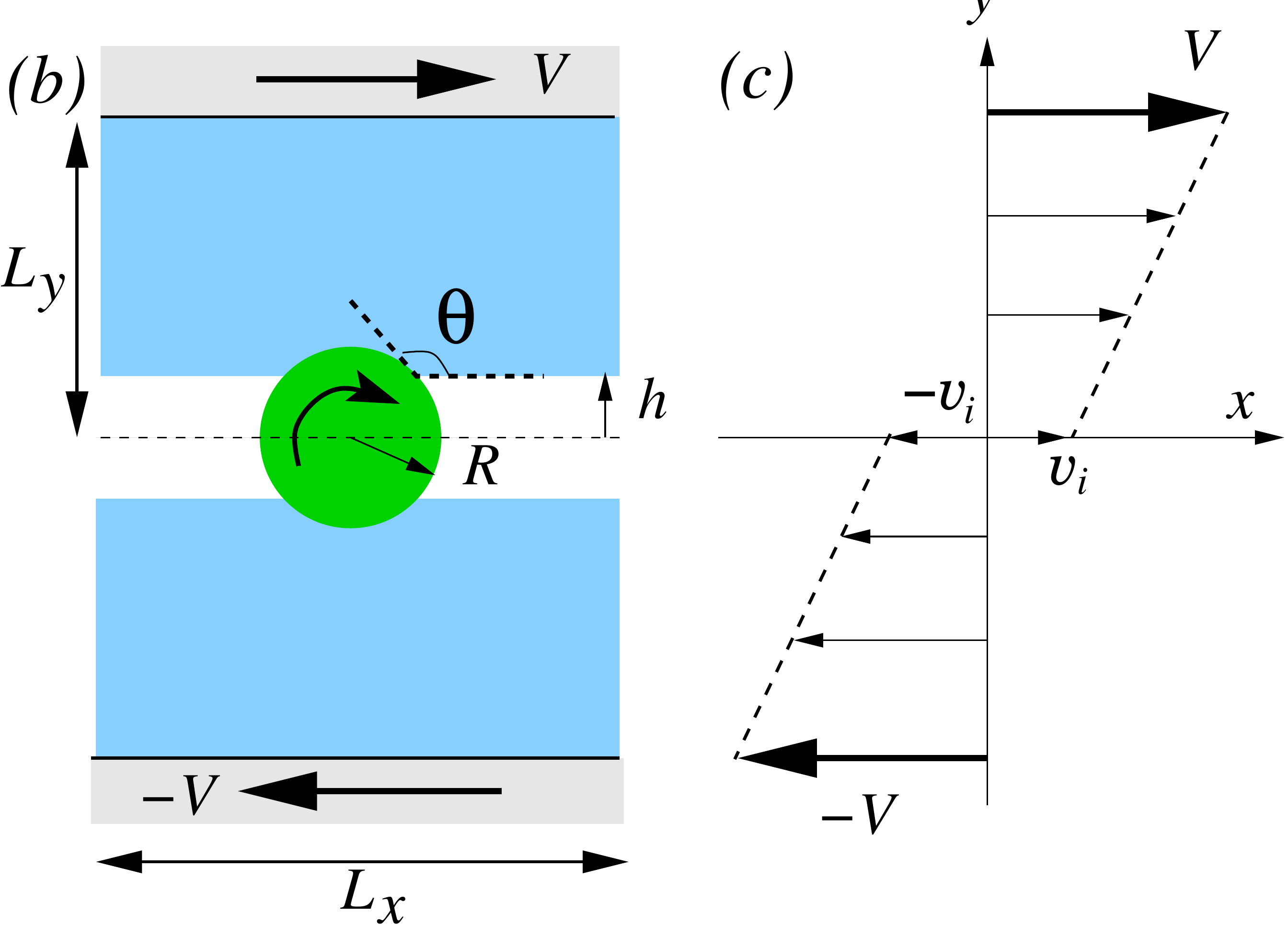}
\caption{{(color online)} Schematics. (a) Using hydrophobic beads as a ball bearing to enhance
liquid/liquid slip. (b) Unit cell of the periodic system analyzed in the text. (c) Expected
macroscopic flow in the interface region.}
\label{fig:concept}
\end{figure}

Starting with the seminal work of Navier in 1823 \cite{Navier1823}, the study of
interfacial slip at liquid-solid interfaces has a long history. 
Slip is usually not observed at macroscopic scales,
but the recent downsizing of hydrodynamic flows in
micro and nanofluidic devices \cite{Stone2004,Bocquet2010} has paved the way for
measurements  at small length scales. These measurements
revealed that slip occurs at the nanometer scale
on flat substrates \cite{CottinBizonne2005,Joly2006,Bocquet2007,Lasne2008,Vinogradova2001}. 
However, nature has produced surfaces such as plant leaves,
with complex topographic structures (bumps, hairs, etc.) on the top of which water
drops can be deposited without collapse in the so-called Cassie-Baxter
(or fakir) state \cite{Cassie1944,Callies2005,Shirtcliffe2010}.
Man-made devices based on this principle 
led to an increase of slip by orders of magnitude,
using e.g. substrates with nanopillar arrays \cite{Choi2006,Joseph2006,Ybert2007,Feuillebois2009,Reyssat2010,Mognetti2010}.
One may therefore wonder whether a similar strategy could
be used to increase slip in liquid-liquid interfaces. 
Indeed, a few studies considering the
possibility of slip at the bare interface between two simple liquids have shown that slip,
if present, was negligible above the molecular scale \cite{Galliero2010,Hu2010}.
Surprisingly, virtually no work has been devoted to optimizing slip at liquid/liquid interfaces.
A large liquid-liquid slip would open new possibilities
in micro and nanofluidics, permitting different liquids in contact
to flow independently with reduced interfacial friction. 
In the following, we show that it is possible to achieve
giant slip between two liquids using a bed of hydrophobic beads,
as depicted in Fig.~\ref{fig:concept}(a).

%
%

Following the principle of usual ball bearings,
we suggest to use an interface where hydrophobic beads maintain a
gas layer between two bulk liquids.  
The stability of this interface is ensured by wetting forces,
instead of being controlled by the contact between the beads and the solid substrate
in  usual ball bearings.
Such a metastable configuration reminds of pillar-based superhydrophobic
surfaces, used to amplify liquid-solid slippage \cite{Choi2006,Joseph2006,Ybert2007,Feuillebois2009,Reyssat2010,Mognetti2010}. 
In both cases the liquid rests mainly on
a gas layer, leading to a low level of friction. However there are two
important differences between these systems: to the advantage of hydrophobic ball bearings,
beads are able to roll, thereby reducing friction at the
liquid/bead interface. However beads will always penetrate
inside the liquid, thereby inducing viscous dissipation, and 
consequently decreasing slippage. 
As seen in  Fig.\,\ref{fig:concept}(a,b), the penetration depth is directly controlled
by the wetting angle $\theta$ of the liquid at the bead surface.
As a consequence, $\theta$ is expected to have a strong influence on the
efficiency of liquid/liquid bearings.
In the following, we start by quantifying analytically the influence of the wetting angle on
liquid/liquid slip in this system. We then confirm the obtained scaling law
by means of Molecular Dynamics (MD) simulations.
Finally, we discuss 
the expected amplitude of slip in experimental systems.


In order to discuss the behavior of liquid-liquid bearings under shear,
we shall consider a simplified geometry with a periodic array of beads.
The unit cell contains a single bead,
as sketched in Fig.\,\ref{fig:concept}(b). Shear is forced
by a velocity difference $2V$ between two parallel walls
separated by a distance $2L_y$. The $x$ and $y$ axes
are indicated in the schematic, and the $z$ axis is 
orthogonal to the plane of the schematic. The system is assumed to be 
periodic along the $x$ and $z$ axes. 
At liquid-solid interfaces, the Navier partial slip
boundary condition (BC) \cite{Navier1823,Bocquet2007}, $\Delta v = b \dot{\gamma}$, relates the
velocity jump at the interface $\Delta v$ to the far-field shear 
rate $\dot{\gamma}$ inside the liquid, where $b$ is the slip length.
The slip length between two liquids with the same viscosity
may be defined using the same equation without ambiguity
(see Supplementary material).
Therefore, one can use the velocity $v^i$
extrapolated at the median plane passing though the center of the bead
as in Fig.~\ref{fig:concept}(c), leading to
\begin{eqnarray}
b={2v^i \over \dot\gamma},
\label{e:b_def}
\end{eqnarray}
where $\dot\gamma=(V-v^i)/L_y$ is the macroscopic shear rate in the
liquid. 

Focusing on small scales hydrodynamics, we 
investigate the limit of small Reynolds numbers
and negligible gravity effects.
{ We consider a steady state. 
In the referential of the bead center of mass,
the solid-liquid and liquid-gas interfaces have null
normal velocity (however, they can exhibit a non-vanishing tangential
velocity). 
In such a fixed geometry, a minimum dissipation principle
is available for Stokes flow (i.e. the limit of vanishing Reynolds numbers) with 
no-slip boundary condition as discussed historically
by  Helmholtz \cite{Helmholtz1868}, Rayleigh \cite{Rayleigh1913}, and
Korteweg \cite{Korteweg1883}.}
This principle is easily extended
to a variable slip length along the boundary~\cite{PierreLouis2013}, and therefore applies
to our system, where the interface is a sum
of portions of liquid-gas interface with infinite slip and
liquid-solid interface with zero or partial slip. 
Since the walls are only an artificial setup to enforce shear, we consider a no-slip BC
at the wall-liquid interface in the following.
However, at the physically relevant liquid-bead interface, we shall
discuss both cases of perfect slip and vanishingly small slip.

Let us start with the case of a perfect slip.
The steady state corresponds to the minimization
of the total viscous dissipation:
\begin{eqnarray}
Q = \frac{\eta}{2} \int \mathrm{d}^3 \mathbf{r}
\left[ \partial_{\mu}v_\nu + \partial_\nu v_\mu \right]^2 ,
\label{e:Q_def}
\end{eqnarray}
where $\mu$, $\nu$ represent $x$, $y$, or $z$,
and the Einstein summation convention is assumed.
Our strategy is to evaluate $Q$ as a function of the interface
velocity $v^i$, and to select the velocity $v^i_*$ for which
$Q$ is minimized.
We therefore need to analyze of the flow field $\mathbf v=(v_x,v_y,v_z)$ present in the system.
The first contribution to $\mathbf v$ is due to the 
direct shear flow ${\bf v}_{D}$ imposed by the boundary conditions.
In the upper part of the system, with $y>0$, we have
\begin{eqnarray}
{\bf v}_{D} = \left(v^i+y\dot\gamma\right)\hat{\bf x}.
\end{eqnarray}
The second contribution is the backflow ${\bf v}_{B}$
caused by the response of the system to the direct flow.  
We assume that the liquid-gas interface remains
flat, so that the backflow is mainly caused by the presence of the bead.
Its main component is along $y$:
\begin{eqnarray}
{\bf v}_{B}=\partial_xy_sv_{Dx}\hat{\bf y},
\end{eqnarray}
where $y_s(x,z)$ is the position of  the 
liquid boundary (liquid-solid plus liquid-gas interfaces).
Since $\theta\rightarrow \pi$,
we have $\partial_{xx}y_s\approx -1/R$ and $y_s\approx R$ above the bead,
leading to 
\begin{eqnarray}
\partial_xv_{By}\approx -{1 \over R} \left(v^i+R\dot\gamma\right).
\label{e:dxvBy}
\end{eqnarray}
Using Eq.\,(\ref{e:Q_def}), the total dissipation now reads:
\begin{eqnarray}
Q={\eta \over 2} \int d^3\mathbf r 
\left[
\dot\gamma-{1 \over R} \left(v^i+R\dot\gamma\right)
\right]^{2}.
\label{e:Q_interm}
\end{eqnarray}
The backflow  penetrates the liquid in a 
domain of volume $\ell^3$, where $\ell\approx 2R(\pi-\theta)$ is the lateral extent
of the solid-liquid contact region. Hence, the
backflow terms in the integrand of Eq.\,(\ref{e:Q_interm})
should be integrated in this volume only.
Using the separation of length scales $\ell\ll R\ll L_y$,
we then find
\begin{eqnarray}
Q = 
\eta {L_xL_z \over L_y}(V-v^i)^2
+ 8 \eta \alpha R (\pi-\theta)^3 (v^i)^2,
\label{e:Q_3D}
\end{eqnarray}
where $\alpha$ is a dimensionless prefactor
accounting for the system geometry.
We may now find the slip velocity $v^i_*$ which minimizes 
the dissipation from the equation $\partial_{v^i}Q_{2D}|_{v^i=v^i_*}=0$.
Finally, using $v^i_*$ in Eq.\,(\ref{e:b_def}), we obtain
\begin{eqnarray}
b=  {1 \over 4\rho R\alpha(\pi-\theta)^3},
\label{eq:b-theta}
\end{eqnarray}
where $\rho=1/(L_xL_z)$ is the bead density at the interface.

The no-slip limit is more subtle,
due to the expected divergence of viscous dissipation at the triple line \cite{DeGennes1985,Huh1971},
as the intrinsic liquid/bead slip length $b_0$ vanishes.
This contribution enters into play as a divergence
of $\partial_{xx}y_s$ at the triple line, replacing
$1/R$ by $(\pi-\theta)/b_0$ in Eq.\,(\ref{e:dxvBy}). 
Assuming that the backflow extends to a distance $b_0$
along the triple-line perimeter $\pi\ell$ then suggests a dissipation
$\sim \eta v^{i2} \pi\ell b_0^2 [(\pi-\theta)/b_0]^2\sim \eta v^{i2}R(\pi-\theta)^3$
as in Eq.(\ref{e:Q_3D}).
A more precise analysis based on the expansion for $\theta\rightarrow \pi$ of the exact two-dimensional
solution of Stokes flow in a wedge \cite{Huh1971} confirms this scaling, but also
indicates the presence of a logarithmic correction.
Indeed, as  $\theta\rightarrow \pi$ the dissipation per unit length of triple line
reads $\sim \eta v^{i2}(\pi-\theta)^2\ln(\ell_0/b_0)$, 
where $\ell_0$ is a macroscopic cutoff. Assuming $\ell_0\sim\ell$, we obtain
a dissipation  $\sim \eta v^{i2}R(\pi-\theta)^3\ln[(\pi-\theta)R/b_0]$, 
from which we compute $b\sim 1/\{R(\pi-\theta)^3\ln[(\pi-\theta)R/b_0]\}$.
We therefore expect 
the scaling behavior in Eq.\,(\ref{eq:b-theta}) to be weakly  affected (i.e. within logarithmic corrections)
by the boundary condition between the liquid and the bead.

{ Note that our analysis relies on
the assumption that $\theta$ does not depend
on the triple-line velocity. This should be achieved
for low capillary numbers\cite{DeGennes1985}. In this limit,
$\theta$ does not depend on $\eta$, and thus
Eq.(\ref{eq:b-theta}) indicates that $b$ should be independent of $\eta$.}

\begin{figure}
\includegraphics[width=\linewidth]{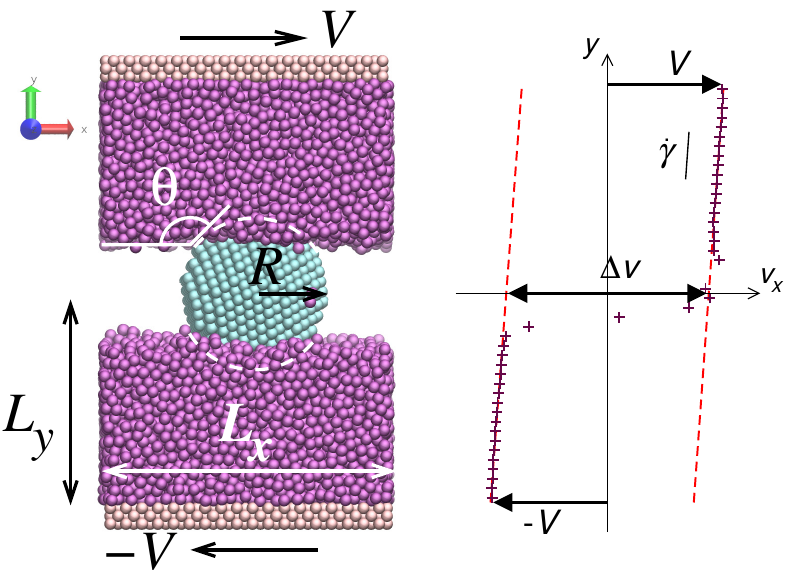}
\caption{{(color online)} Snapshot of a typical molecular dynamics system 
(created using VMD~\cite{VMD}), with the
associated steady-state velocity profile on the right.}
\label{fig:MD}
\end{figure}

In order to test the predictions of Eq.~\eqref{eq:b-theta}, we performed
molecular dynamics (MD) simulations, using the LAMMPS package \cite{LAMMPS}.
We consider once again a periodic array of beads, 
the unit cell of which is presented in  Fig.~\ref{fig:MD}.
%
%
All atoms interacted via a Lennard-Jones (LJ) pair potential: $V_{ij} = 4
\varepsilon_{ij} [(\sigma/r)^{12}-(\sigma/r)^{6}]$, where
$i$ and $j$ are the types of interacting atoms, and $r$ 
and $\varepsilon_{ij}$ are respectively their distance and interaction energy.
The zero-potential distance $\sigma$ is the
same for all atom types. In addition, we chose the fluid-fluid interaction energy as the
reference energy $\varepsilon_{ff} = \varepsilon$.
In the following, we will use LJ units of energy
$\varepsilon$, distance $\sigma$, and time $\tau = \sigma
\sqrt{m/\varepsilon}$, where $m$ is the atomic mass.
In order to increase the liquid/vapor
surface tension and to decrease the vapor density as compared to a simple
LJ liquid, we used a liquid of LJ dimers \cite{Galliero2010a}, 
i.e. pairs of LJ atoms bonded by a harmonic potential. 
We used a
dissipative particle dynamics (DPD) thermostat to maintain the liquid
atoms at a temperature $k_\mathrm{B}T = 0.73\varepsilon$ while
preserving hydrodynamics \cite{Groot1997}.
The bead was modeled as a rigid
spherical shell of radius $R = 7.07 \sigma$ (with a thickness
larger than the LJ cutoff $2.5\sigma$), cut inside a fcc crystal with density $1/\sigma^{3}$.
The confining walls were made of three (010) layers of frozen atoms on a
fcc lattice, with the same density $1/\sigma^{3}$. 
The contact angle is controlled by the fluid-solid interaction energy
in the LJ potential. We used $\varepsilon_{fw} = 0.7 \varepsilon$
between fluid and wall atoms, so as to ensure a small contact
angle, and consequently a no-slip BC \cite{Huang2008b}.
In contrast, the fluid-bead interaction energy
$\varepsilon_{fb}$ was tuned in order to change the contact angle $\theta$. The
latter was then computed from \textit{in situ} measurements of the
vapor gap width.
Finally, the distance between the top of the bead and the
walls was $\approx2 R$ in all simulations.
This was large enough to remove in practice the influence of the
confining walls, and to allow for a linear average velocity
profile in the liquid which enables the measurement of the shear rate $\dot\gamma$.
After a period of equilibration, the system was sheared by imposing
opposite wall velocities $\pm V$ along $x$. 
All along the simulation, both walls were used as pistons
to impose a vanishing pressure inside the system. 

\begin{figure}
\includegraphics[width=0.9\linewidth]{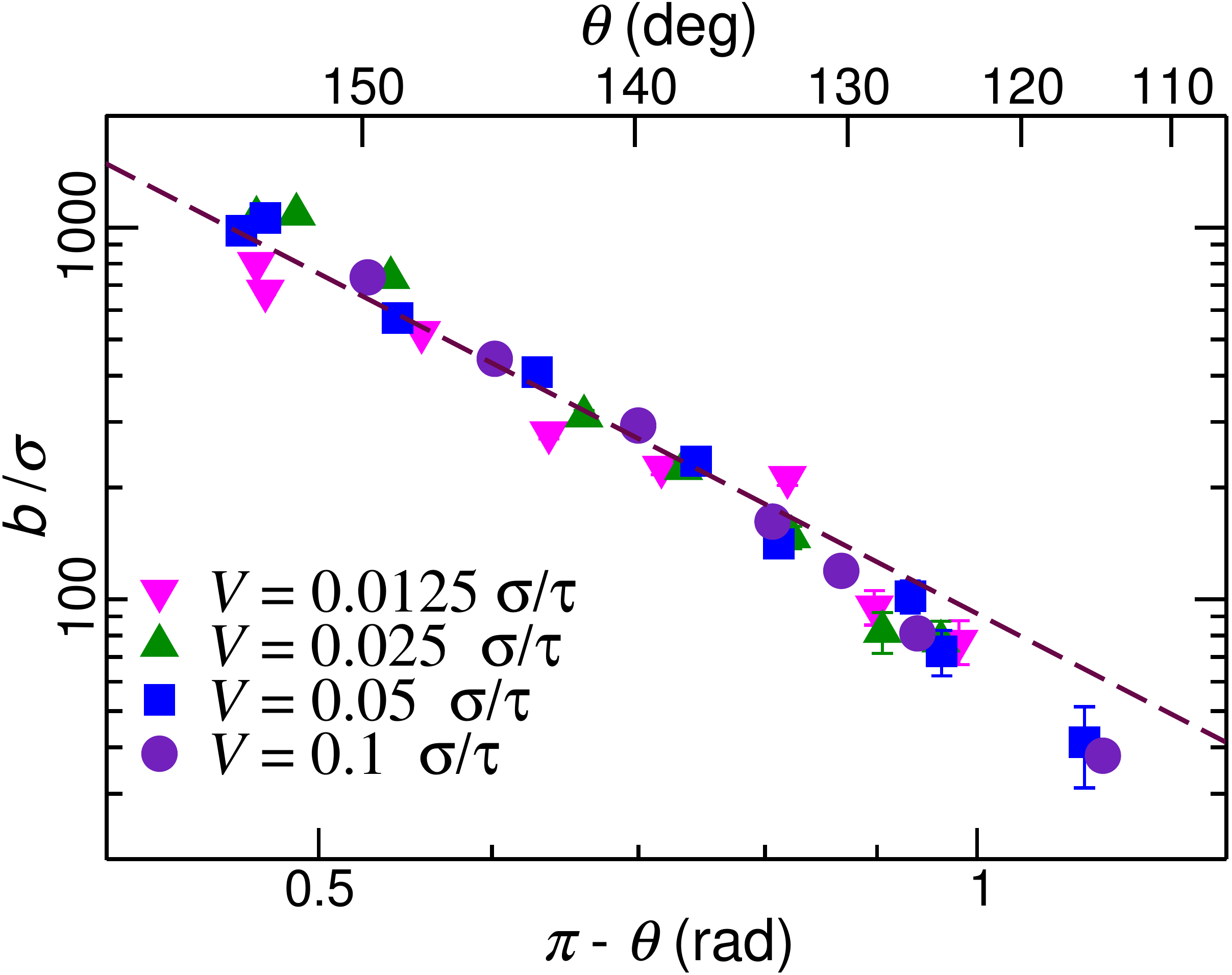}
\caption{{(color online)} Slip length $b$ as a function of the measured wetting angle
  $\theta$, using the dynamic definition of the slip length. 
  Data set keys indicate the shear velocity $V$, in LJ units ($\sigma/\tau$). All data
  sets collapse on a master curve $b \propto (\pi -
\theta)^{-3}$ (dashed line).}
\label{fig:result}
\end{figure}

To measure $b$, we first used the kinematic definition,
Eq.\,(\ref{e:b_def}).
In order to extract $\Delta v$ and $\dot{\gamma}$, we fitted the numerical velocity
profiles in the linear regions between the bead and the walls, as
shown in Fig.\,\ref{fig:MD}. 
However, for large contact angles, the shear rate $\dot{\gamma}$ becomes increasingly
small, and its measurement in the fluid is inaccurate. 
We therefore resort to a different definition of the slip length based on 
a dynamic interpretation \cite{Falk2010}. On the one hand, the bulk
viscous shear stress is $\eta \dot{\gamma}$, where $\eta$ is the
liquid viscosity.
On the other hand, the interface shear stress
equals the interfacial friction force $f$
per unit area, which is proportional to the velocity jump: $f = \lambda\Delta v$. 
This latter relation defines the interfacial friction coefficient
$\lambda$. Equating the bulk and interface shear
stresses, one recovers the partial slip BC with $b = \eta / \lambda$. 
Therefore, the slip length can
also be computed as the ratio between the bulk liquid viscosity and
the interfacial friction coefficient. 
The liquid viscosity $\eta\approx 5.52$ LJ units, was determined under the same thermodynamic
conditions in independent shear flow simulations. The friction
coefficient $\lambda$ was computed as the ratio of the measured shear
force per unit area $f$ and the measured velocity jump
$\Delta v$.
For the analysis of the  non-trivial $\theta$-dependence of $b$, 
we used a fixed box size $L_x = L_z= 4R$, { corresponding to 
$\rho=1/(16R^2)$}. 
In Fig.\,\ref{fig:result}, the resulting slip length is plotted as a 
function of $\theta$. 
%
%
Using the dynamic method, all data points 
collapse on a single master curve $b \propto (\pi -
\theta)^{-3}$, as predicted by Eq.\,(\ref{eq:b-theta}).
The fit to Eq.\,(\ref{eq:b-theta}) provides a value for
the dimensionless prefactor $\alpha = 0.31 { \pm 0.02}$.
The results of the kinematic method are similar to those of the 
dynamic method. However, as anticipated above, the kinematic method is less 
precise at small shear rates, leading to larger scatter of the data. The comparison between 
the results of the two methods is provided in Supplementary Material.
{ We have also checked that $b\sim 1/\rho$, as predicted by
Eq.~(\ref{eq:b-theta}).
Deviations from this latter scaling are observed when hydrodynamic
interactions come into play at high bead-density and
when $\pi-\theta$ is not small. Details are discussed
in Supplementary Material.}

A recurrent issue in MD simulations is the need to use a very
large forcing in order to extract the system response out of thermal
noise. Such large driving forces could lead to a nonlinear response,
questioning the description of the interface behavior by means of a slip length.
Our shear rates were $\sim 10^{7}$ to $10^{9}$\,Hz, which is very large 
as compared to experimentally relevant shear rates, typically smaller than $10^5$\,Hz.
As shown in Fig.~\ref{fig:result}, we varied $V$ over almost one order of
magnitude and checked that the measured slip lengths remained
unaffected. We may therefore conclude that the system response is 
linear, and can be safely extrapolated to the much lower experimental shear rates. 
{ In addition, it is clear from the adequacy
between Eq.(\ref{eq:b-theta}) and MD simulations that 
thermal fluctuations play a negligible role
in the emergent macroscopic behavior, 
at least in the explored parameter range.}

Finally, note that we do not control the slip at the liquid-bead interface in our MD simulations.
It is known that both the contact angle and the curvature of the solid surface
may have a strong impact on the liquid-bead slip\cite{Huang2008b,Einzel1990}.
However, as discussed above, the variation of the liquid-bead slip length $b_0$
is not crucial for our main result Eq.\,(\ref{eq:b-theta}), 
and should lead to minor logarithmic corrections,
the discussion of which are beyond the scope of the present Letter.


Interfaces similar to that of Fig.\ref{fig:concept}(a) have actually
already been observed experimentally using liquid marbles
floating on liquid substrates\cite{Quere2002,Bormashenko2012}.
However, to our knowledge no analysis of slip has been performed
for these interfaces.
An important way
to reduce slip in these systems is to decrease the bead density, which directly enters into
the expression of $b$ in Eq.(\ref{eq:b-theta}). In addition, lowering the density may help
to prevent contact and friction between the beads, which is expected have negative consequences on slip.
One possible way to control the density is to use electrostatic repulsion between the beads
along the same lines as the in formation of colloidal Wigner crystals at water-oil interfaces~\cite{Irvine2010}.
Let us assume that one could build such a system,
with typical densities $\rho$ varying from $0.01R^{-2}$ to $0.1R^{-2}$.
Assuming $0.03R^{-2}$ (corresponding to a distance of about two diameters
between the beads), Eq.(\ref{eq:b-theta}) predicts $b\approx 20R$ with $\theta=120\,^\circ$, 
and $b\approx 200R$ with  $\theta=150\,^\circ$ (using
e.g. superhydrophobic beads). With a bead radius $R=10\,\mu$m,
$b$ would therefore be on the scale of millimeters for large contact angles
($b\approx 0.2$ and $2$~mm for $\theta=120$ and $150\,^\circ$ respectively).

In our opinion, a strong limitation of this giant slip is the stability
of the interface under pressure variations. 
Two processes are involved. First, the
Laplace-pressure-induced curvature of the liquid-gas interface
imposes an additional backflow, increasing dissipation. 
This effect is similar to the effect of pressure
in pillar-based superhydrophobic surfaces~\cite{Steinberger2007,Biben2008}. 
Defining the curvature
$\kappa$ of the liquid-gas interface, and assuming that the related backflow
extends in the liquid up to a distance similar to the inter-bead distance, we obtain
an additional dissipation $\sim \eta \rho^{-3/2}\kappa^2(v^i)^2$ in Eq.\,(\ref{e:Q_3D}).
{ 
Recalling the Laplace law $\kappa=\Delta P/2\gamma$, where $\Delta P$ is the pressure
variation and $\gamma$ is the surface tension, we obtain
a correction of the slip length 
$b^{-1}=b^{-1}|_{\Delta P=0}+\alpha'\rho^{-1/2}(\Delta P)^2/4\gamma^2$
with $\alpha'$ a constant of order one, showing that $b$ decreases
when increasing $|\Delta P|$. 
The correction is negligible when}
$|\Delta P|< 2\gamma \rho^{1/4}b^{-1/2}$, where the zero-pressure slip length
$b$ is given by Eq.\,(\ref{eq:b-theta}). This condition becomes stringent
when $b$ is large, and for  $\theta=150\,^\circ$ with $R=10\,\mu$m,
one obtains $|\Delta P|<8$\,mbar. The second limitation related to the pressure
is the possibility of contact between the two opposite liquid-gas interfaces,
leading to the collapse of the system.
Balancing the typical interface displacement $\kappa\rho^{-1}$ with the 
typical average separation $\sim R$, the collapse will be avoided when
$\Delta P< 2\gamma R \rho$. Using 
$R=10\,\mu$m and $\rho=0.03R^{-2}$, one finds again $P<4$\,mbar. 
The stability can be enhanced, at the cost of a smaller slip length,
by using higher densities and smaller beads. For instance, with $R=0.1\,\mu$m and $\rho=0.1R^{-2}$,
one finds $P<1$\,bar, but $b \approx 6\,\mu$m for $\theta=150\,^\circ$.


In conclusion, giant slip can be obtained
between two liquids using hydrophobic ball bearings. 
We hope that this work will motivate
experimental investigations of the dynamic properties of such interfaces.
Liquid-liquid bearings open new pathways
for micro- and nano-fluidics research. One major direction
could be to build fluidic devices without walls,
where different liquids could flow with regard to each
other while maintaining an extremely low level of friction, and
preventing mixing by diffusion between the different channels.

\begin{acknowledgments}
OPL wishes to thank J. Yeomans and H. Kusumaatmaja for discussions.
\end{acknowledgments}

\bibliography{laurent,laurent1,laurent2,opl_ref}

\end{document}